\documentclass[aps, nofootinbib, preprint, showpacs]{revtex4}

\usepackage{amssymb}

\usepackage{epic}    
\usepackage{eepic}

\newcommand {\rem}  [1] {}
\newcommand {\mtext}[1] {\quad \mbox{#1} \quad}

\newcommand {\nn}   {\nonumber}
\newcommand {\be}   {\begin{equation}}
\newcommand {\ee}   {\end{equation}}
\newcommand {\bea}  {\begin{eqnarray}}
\newcommand {\eea}  {\end{eqnarray}}
\newcommand {\bit}  {\begin{itemize}}
\newcommand {\eit}  {\end{itemize}}

\newcommand {\Sch}  {Schwarz\-schild}
\newcommand {\BD}   {Brans--Dicke}

\newcommand {\Ray}  {Ray\-chau\-dhu\-ri}
\newcommand {\RN}   {Reiss\-ner--Nord\-str\"om}
\newcommand {\EF}   {Edd\-ing\-ton--Fin\-kel\-stein}
\newcommand {\ArPi} {Ar\-men\-d\'a\-riz--Pi\-c\'on}
\newcommand {\MTW}  {Mis\-ner, Thor\-ne and Whee\-ler}
\newcommand {\GHS}  {Gar\-fink\-le, Horo\-witz and Stro\-min\-ger}
\newcommand {\EFP}  {Es\-po\-si\-to--Fa\-r\`e\-se and Po\-lar\-ski}
\newcommand {\MED}  {Max\-well--Ein\-stein--Di\-la\-ton}
\newcommand {\KG}   {Klein--Gor\-don}
\newcommand {\SSS}  {sta\-tic sphe\-ri\-cal\-ly sym\-met\-ric}

\newcommand {\Ef}   {\mbox{E--frame}}
\newcommand {\Sf}   {\mbox{S--frame}}
\newcommand {\Gf}   {\mbox{G--frame}}
\newcommand {\Af}   {\mbox{A--frame}}

\begin{document}

\preprint{\vbox to 30 pt{
\hbox{UWThPh--2002--24}
\hbox{gr-qc/0209002}
\vfil}}

\title{Geometric Dilaton Gravity and Smooth Charged Wormholes} 

\author{Wolfgang Graf}
 
\affiliation {Institut f\"ur theoretische Physik der Universit\"at Wien, \\
Boltzmanngasse 5, A--1090 Wien, Austria}

\email {wolfgang.graf@ap.univie.ac.at}


\date{\today}

\begin{abstract}
A particular type of coupling of the dilaton field to the metric
is shown to admit a simple geometric interpretation in terms of a volume element density
independent from the metric. 
For dimension $n=4$ two families of either magnetically or electrically charged 
\SSS~solutions to the corresponding Maxwell--Einstein--Dilaton field equations are derived.
Whereas the metrics of the ``magnetic'' spacetimes are smooth, asymptotically flat
and have the topology of a wormhole, the ``electric'' metrics 
behave similarly as the singular and geodesically incomplete classical \RN~metrics.
At the price of losing the simple geometric interpretation,
a closely related ``alternative'' dilaton coupling can nevertheless be defined,
admitting as solutions smooth ``electric'' metrics.
\end{abstract}

\maketitle

\section{Introduction}

Einstein's classical theory of gravity, based on a metric, 
has so far passed all experimental tests concerning the motion of bodies or the
deflection of light, with ever increasing precision (cmp.~Will~\cite{Wil01}).
But there are some disturbing limitations of a more 
formal nature, like generic singularities, when not allowing ``exotic matter''.
This is most evident in the gravitational collapse of stars
and in the early phases of the Universe, notably at the ``big bang singularity''.
In fact, it seems that we have come to live with this 
unsatisfactory situation as a necessary consequence
of a classical description.
These problems can be attributed either to
Einstein's theory itself or to the inadequacy of the matter model used --- or to both.
In the first case, singularities seem to be a generic feature
(somewhat attenuated by the expectation that they will  
in general be hidden behind an event horizon),
unless some other ``pathologies'' are accepted, like exotic matter or closed timelike lines.
In the other case, for example the apparent accelerating cosmic expansion
has led to a generalization involving a ``variable cosmological constant'' $\Lambda$
(``Quintessence'', ``k--essence''). Previously, already a generalization based on a 
``variable gravitational constant'' $G$ has been considered, well--known as \BD~theory.
Also, in order that inflation works, some specific modifications must be made.
In all these generalizations, an additional real scalar field,
serving the particular purpose, plays a fundamental role.
Save in exceptional cases, the singularity problem remains.
It is however generally expected, that a future reconciliation of
Gravity and Quantum Theory will lead to a unique theory
without the above mentioned problems.
Attempts in this direction can be seen
in the currently extensively studied ``string--inspired cosmologies'',
which are based on some low energy limit of String Theory
and an appropriate reduction to dimension $n=4$.
There a scalar field multiplying (as an exponential) 
the Ricci--scalar and/or the metric  plays a prominent role,
which is interpreted physically as the dilaton field.
A big advantage of these approaches is the fact
that equations closely related to Einstein's are a necessary consequence.
\medskip

Our aim is to show that already on a classical level,
by properly including the dilaton scalar,
some of the singularity problems can be avoided.
Firmly based on Standard Differential Geometry and only
loosely inspired by the dilaton scalar of String Theory,
we formulate a theory of Dilaton Gravity.
As criteria for the soundness of our approach, 
both the formulation of a minimal coupling scheme and
the existence of nontrivial geodesically complete and asymptotically flat
solutions of the corresponding field equations is taken.
\medskip

We will proceed as follows.
In section~\ref{coupling} a particular form of the coupling of a scalar field $\phi$
(called ``dilaton'') to the metric tensor $g_{ik}$ of a spacetime is proposed,
which admits a straightforward and unique geometric interpretation in terms of an 
independent {\em Volume Element Density}.
A dilatonic coupling scheme is formulated, in order to accomodate additional nongeometric fields.
In section~\ref{equations}, the field equations corresponding to a \MED~Lagrangian are derived,
where also their ``Einstein form'' is given.
In section~\ref{solutions} a family of magnetically charged
\SSS~solutions is derived, closely related to the well--known ``string--inspired'' 
charged black hole solutions of \GHS~\cite{GHS91}.
These metrics are shown to be geodesically complete (in fact, smooth) and asymptotically flat,
each of them containing a {\em wormhole}, with no exotic matter being involved.
The corresponding family of electrically charged solutions consist of singular metrics
which have either an event horizon or exhibit a naked singularity.
An ``alternative'' nongeometric coupling is shown to be however possible,
admitting a family of smooth electrically charged solutions.
In section~\ref{interpretation}, for the proposed geometric coupling
the Equivalence Principle is shown to be fulfilled, in the sense that
uncharged point particles still move on geodesics.
In order to derive expressions for the mass and some other significant parameters,
the paramatrized post--Newtonian approximation is invoked.
In terms of the basic parameters $m$, $\beta$ and $\gamma$, 
it is shown that the derived solutions must be considered as viable with respect
to present--day astronomical empirical data.
For the realm of elementary particles,
it is shown that for the smooth magnetic and electric wormhole solutions,
significant effects could be expected at distances of
roughly the order of the classical electron radius.
For the wormhole solutions it is shown that their mass
parameter has zero as lower bound.
Also an explanation of the ``repulsive'' character
of the dilaton involved in the wormhole solutions is given.
The concluding section~\ref{conclusions} formulates the main conclusions
and points to some important open questions.

\section{Geometric Dilaton Coupling}
\label{coupling}

\subsection{Volume Geometry; Hodge Duality}

As well--known (cmp.~e.g.~Abraham, Marsden and Ratiu~\cite{AMR88} 
for the closely related concept of a {\em Volume Manifold}), 
a Volume Element Density (VED) is geometrically 
a nondegenerate smooth $n$--form density vol, that is,
under orientation--preserving coordinate transformations it behaves as
a conventional $n$--form, whereas under orientation--reversion
it gets an extra factor $-1$.
Such a VED already allows to invariantly express the divergence div $v$ 
of a vector field $v$
as $(\mbox{div}~v)\;\mbox{vol} := d\;(v \cdot \mbox{vol})$,
where the dot denotes the contraction of a differential form by a vector field.
This definition is well--known in Hamiltonian Mechanics, where
it plays a major role. Also the Gauss integral theorem can already be formulated.
Note that no metric has been involved so far.

It makes sense to speak of a ``positive'' VED,
and any two such VEDs differ only by a positive function $\lambda$.
Assuming that we also have a nondegenerate metric $g_{ik}$,
we can therefore always write for a general VED,
vol = $|g|^{1/2}\;e^\phi\;dx^1 \wedge \cdots \wedge dx^n$,
where we have conveniently set $\lambda = e^\phi$, with some scalar function $\phi$.\footnote
{as is common practice in the physics literature, we will  denote the corresponding
coefficient of the $n$--form $dx^1 \wedge \cdots \wedge dx^n$ 
as ``scalar density'' --- e.g., the Lagrangian~$\mathcal L$}
This functional form ensures positivity of $\lambda$, when $\phi$ is continuous.
Of course we could also have chosen any other smooth monotone positive function of $\phi$,
but this would not introduce anything new, 
as effectively only the ``dilaton factor'' $e^\phi$ matters.
The scalar field $\phi$ thus represents an 
{\em essentially unique new geometrical degree of freedom}.

Its occurrence in form of the factor $e^\phi$
strongly reminds of the dilaton factor
appearing in the reduced Lagrangians for the 
Low Energy Limit of String Theory (LELoST).\footnote
{although there the equivalent factor $e^{-2\phi}$ seems to be more natural}
\medskip

For a volume manifold with a nondegenerate metric, the notion of the
{\em Hodge dual} of a differential form has to be sligthly generalized.
Recall that the dual $\star F$ of a plain $p$--form $F$ is the result of the following
construction, given with respect to some coordinate basis:
\bea
F_{i_1\cdots i_p} &\to& F^{j_1\cdots j_p} := g^{i_1j_j}\cdots g^{i_pj_p}\;F_{i_1\cdots i_p} \nn \\
                  &\to& \star F_{j_{p+1}\cdots j_n} := \mbox{vol}_{j_1\cdots j_p j_{p+1}\cdots j_n}\;F^{j_1\cdots j_p} \\
                  &\equiv& |g|^{1/2}\;e^\phi\;\varepsilon_{j_1\cdots j_p j_{p+1}\cdots j_n}\;F^{j_1\cdots j_p}, \nn
\eea
where $\varepsilon$ denotes the permutation symbol.
The plain $p$--form $F$ is thus mapped to a  $(n-p)$--form density $\star F$.
As this map is one--to--one, it can be inverted to map an $p$--form $H$ density to a plain $(n-p)$--form $\star^{-1} H$.
Unfortunately, the nomenclature of the two different types of differential forms is far from standard\footnote
{also the following designations are common: pseudoforms, twisted forms, Weyl--tensors, oriented tensors}
and so we will adhere to de Rham~\cite{Rha84}, denoting the forms we called ``plain'' with {\em even type}
and the ``form densities'' with forms of {\em odd type}.

Consequently, we now define the generalized Hodge dual of any form $F$ (even or odd) as
\be
\ast F = \left\{ \begin{array}{rl}  \star F, &F \ \mbox{even}, \ \\ 
                                    \star^{-1} F, &F \ \mbox{odd} . \end{array} \right.
\ee
This definition makes the duality operator trivially idempotent, $\ast^2 F = F$.\footnote
{note that when distinguishing even and odd type forms, $\star^2$ does not make sense}
\smallskip

Alternatively, we could also define the Hodge duality $F \to \ast F$ as the {\em unique isomorphism}
from the vector space of even (odd) $p$--forms to the ``dual'' vector space 
of odd (even) $(n-p)$--forms, whose restriction to even forms gives
\bea
F \wedge \ast G &=&  (F,G)\;\mbox{vol}, \qquad F,G \ \mbox{even}. 
\eea
Here the round bracket denotes the scalar product of forms
based on the Riemann metric.
As a consequence, we have for odd forms the corresponding relation
\bea
\ast F \wedge G &=& (\ast F,\ast G)\;\mbox{vol}, \quad F,G \ \mbox{odd}.
\eea

Based on the Hodge duality for differential forms, the operators for the {\em divergence} $\delta$ 
and the {\em Laplacian} $\Delta$ for differential forms
can now be defined as $\delta := \ast d \ast$ and $\Delta := d\delta + \delta d$,
where $d$ denotes the operator of {\em exterior derivative}, which is valid
for forms of any even/odd type.\footnote
{the more precise definition of $\delta$ by de Rham introduces an extra factor $-1$, 
depending on the dimensionality of the manifold and the signature of the metric}
\smallskip

Why this insistence on differential forms?
The main reason is that the Lagrangian scalar density is geometrically more properly understood as an odd form 
of maximal degree $n$, the energy momentum tensor being
a covector--valued odd $(n-1)$--form.
Of course, the electromagnetic Maxwell field, to be extensively used later together with its Hodge dual,
is to be understood as an even 2--form.
Also we need the divergence of a vector field and of a two--form, as well as the Laplacian of a scalar field.

\subsection{General Dilaton Coupling}

Be it from five--dimensional Klein--Kaluza reduction
(for short, KK--reduction),
or from a LELoST compactified to $n=4$,
the Lagrangians have the generic form\footnote
{we use the conventions of \MTW~\cite{MTW73} throughout, 
the squares denoting the conventional metric scalar product}
\be 
\mathcal L = |g|^{1/2}\;e^{\alpha\phi}\;\big( R - \beta\; (\nabla \phi)^2 - e^{\gamma\phi}\;F^2 \big).
\ee 
The constant parameters~$\alpha,\:\beta,\:\gamma$~depend on the particular 
higher--dimensional base theory and the chosen reduction.
Of course, from a LELoST many more scalar fields  and antisymmetric tensor
fields of different degrees (``moduli'') will appear, 
but we keep only the dilaton scalar $\phi$ and a rank--2 antisymmetric tensor 
field $F$, later to be interpreted as the Maxwell field two--form.\footnote
{sometimes ``potential'' terms $V(\phi)$ also appear or are introduced ``by hand''}
$R$ denotes the omnipresent Riemann curvature scalar of General Relativity.
We will refer to this class of Lagrangians as \MED~Lagrangians (MED).

Often, more general Lagrangians are studied, containing free functions
(cmp.~\EFP~\cite{EsP00}, and references therein), 
but our choice is already general enough to cover
the most important applications as special cases.
In particular, the class of Bergmann--Wagoner Lagrangians extensively studied since about 30~years
by Bronnikov et al.~(cmp.~\cite{BrG02}) should be mentioned,
having the general form
$\mathcal L = |g|^{1/2}\;( f(\phi)\:R + h(\phi)\:(\nabla \phi)^2 - F^2)$ 
(in particular, with $h=1$ and $f = 1 - \xi\:\phi^2$, where $\xi$ a constant parameter).
Although diverse charged wormhole solutions have been obtained,
they all violate some of the energy conditions.
Note that the class of MED Lagrangians considered in our work 
essentially differs from Bronnikov's class,
which completely excludes ``string--inspired'' Lagrangians.

As convenient for dimensionally reduced Lagrangians, the relativistic gravitational constant $\kappa$
(as well as any factor $1/2$) is assumed to be absorbed into $F^2$.
For example, the KK--reduction leads to $\alpha=1,\;\beta=0,\;\gamma=2$,
whereas a typical LELoST reduction has $\alpha=-2,\;\beta=-4,\;\gamma=0$.
With a scalar defined by $\Phi := e^\phi$, 
we can also deduce the scalar--tensor \BD~Lagrangian from this form,  
with parameters $\alpha=1,\;\beta=\omega \neq 3/2,\;\gamma=-1$.
Similarly, for $\beta = 3$ we get a conformal scalar coupling.

However, all these MED Lagrangians can be transformed 
modulo a trivial divergence by a Weyl conformal transformation
$g'_{ik} = g_{ik}\;e^{\alpha\phi}$ into a so called {\em Einstein--frame},
characterized by $\alpha'=0$. For dimension $n=4$, this results in 
$\beta'=\beta + 3/2\;\alpha^2,\;\gamma'=\gamma+\alpha$.
In such a frame, formally the conventional Einstein Lagrangian is obtained,
with a massless \KG~(KG) field $\phi$ and a Maxwell field $F$ gravitationally coupled
with an effective coupling ``constant'' $e^{(\alpha+\gamma)\phi}$.
The scalar field $\phi$ is ghost--free, i.e., {\em non--exotic}, if and only if $\beta' \geq 0$.

More generally, for any dimension $n$, a dilaton--based conformal transformation of the metric, 
$g'_{ik} = g_{ik}\;e^{\lambda\phi}$ ($\lambda$ a constant parameter),
leads to the following transformation behaviour of the parameters of the MED Lagrangian density
\bea
\alpha &\to& \alpha' = \alpha - \lambda \nn \\
\beta  &\to& \beta'  = \beta + (n-1)(n-2)/4\:(\alpha^2 - {\alpha'}^2) \\
\gamma &\to& \gamma' = \gamma + \lambda. \nn
\eea

Although the causal structure is not altered as long as $\phi$ is continuous,\footnote
{which is however not in general the case}
some basic metric--based relations (e.g.~length and ``straightness'') are not.
For the physical interpretation therefore some conformal frame has to be 
taken as the fundamental one, characterized by a particular form of the Lagrangian density. 
This will depend on the particular coupling
to other fields, in particular, to point masses.
However for ``string--inspired'' dilaton theories most authors agree that
an Einstein--frame should be taken,
primarily justified by the availability of the familiar interpretatory 
apparatus of the classical Einstein gravity.\footnote
{cmp.~e.g.~Gasperini and Veneziano~\cite{GaV02}}

\subsection{Geometric Dilatonic Coupling and Minimal Coupling Scheme}

Let us now introduce a particularly simple coupling,
characterized by the parameters $\alpha=1,\;\beta=0,\;\gamma=0$.
Evidently, the dilaton enters the corresponding MED Lagrangian in a mode which can be interpreted
geometrically in terms of a general metric--based Volume Element Density,
as previously described. 
The Lagrangian simply becomes  $\mathcal L = |g|^{1/2}\;e^{\phi}\;( R - F^2 )$.
Let us call such a coupling a {\em Geometric Dilaton Coupling} (GDC)
and the corresponding conformal frame a \Gf~(G for ``geometric'').%\footnote
%{in the context of two dimensional string--related theories
%sometimes also denoted by ``twiddle frame''}

The particularly simple form of the GDC suggests the following 
{\em Minimal Geometric Dilatonic Coupling Scheme} (MGDCS):
assuming we have already a Lagrangian density without dilaton
and satisfying the prerequisites of general covariance,
$\mathcal L_0 = |g|^{1/2}\;L$,
we get a GDC by just correcting any occurence of the Riemann volume element density 
$|g|^{1/2}$ by the dilaton--factor $e^{\phi}$:
$\mathcal L := \mathcal L_0\;e^\phi = |g|^{1/2}\;e^{\phi}\;L$ ---
even when it explicitly occurs inside $L$.

Taking the standard Maxwell--Einstein Lagran\-gian as the prototypical example,
we get as result of the MGDCS the GDC Lagrangian:
\be
\mbox{MGCDS:} \quad \mathcal L = |g|^{1/2}\:( R - 1/2\;F^2 ) \longmapsto
\mathcal L = |g|^{1/2}\;e^\phi\;( R - 1/2\;F^2 ).
\ee

Similarly, a massive point particle could also be incorporated, to give
$\mathcal L = |g|^{1/2}\;e^\phi\;R - m\;(\dot x^2)^{1/2}\;\delta_T$,
where the Dirac delta--distribution is supported by the world--line $T$ of the particle,
given by $T: \tau \mapsto x^i(\tau)$.
Note that the mass term does not acquire a dilaton factor, 
as it does not contain the factor $|\det g|^{1/2}$.

\section{GDC Field Equations}
\label{equations}

The Field Equations derived from the geometrically coupled
Maxwell--Einstein--Dilaton Lagrangian 
$\mathcal L = |g|^{1/2}e^{\phi}\;(R - 1/2\;F^2)$ are,
up to a factor $|g|^{1/2}e^{\phi}$,
\bea
\label{PFE_1}
G_{ik} - \sigma^2\:\Theta'_{ik} - \sigma\:\Theta''_{ik} &=& M_{ik}  \\
\label{PFE_2}
R &=& 1/2\;F^2  \\
\label{PFE_3}
0 &=& \mbox{div}\;F,  \\
\mtext{where} M_{ik} &:=& F_{ir} F_{ks} \; g^{rs} - 1/4\; F^2 \; g_{ik},  \\
\Theta''_{ik} &:=& \nabla_i\nabla_k\phi - \nabla^2\phi\;g_{ik},  \\
\Theta'_{ik}  &:=& \nabla_i\phi\nabla_k\phi - (\nabla\phi)^2\;g_{ik}, \\
\mtext{and} \quad \ \sigma &:=& (n-2)/2.  
\eea
Here div $F$ denotes the dilaton--generalized divergence of a 2--form, which in a coordinate base
can also be written as 
\be 
(\mbox{div}\;F)^i := |g|^{-1/2}e^{-\phi}\;\partial_j\;(|g|^{1/2}e^{\phi}\;F^{ij}) \equiv \nabla_j\;F^{ij} + \phi_j\;F^{ij}.
\ee 
Also we have assumed that $F$ has a gauge potential, $F=dA$, 
with respect to which the corresponding variation is performed.
Combining the equation (\ref{PFE_2}) with the trace of equation (\ref{PFE_1}) 
and assuming $n \geq 2$,
we get an explicit {\em dilaton equation} for $\Phi := e^{\sigma\phi}$,
\be
\nabla^2 \Phi - 1/(n-1)\:R\:\Phi = 0.
\ee
{\bf Remarks} \\
i)~neither the dilaton scalar $\phi$ nor the  dilaton factor $e^{\phi}$ do explicitly
appear in the primary field equations (\ref{PFE_1}) -- (\ref{PFE_3}), except through their derivatives --- 
this invisibility of the ``effective gravitational coupling constant'' $e^{\phi}$ 
underlines the geometric character of the theory. \\
ii)~the dilaton scalar does {\em not} couple to the trace (which here vanishes for $n=4$) 
of the energy momentum tensor as in the \BD~theory, but to the independent scalar $F^2$. \\
iii)~the tensor $\Theta'$ has almost the form of the energy momentum tensor
for a massless \KG~field, except for a factor $-1$ instead of $-1/2$. \\
iv)~the dilaton equation is purely geometric, as it does not contain the Maxwell field ---
for dimension $n=6$, it reduces to the {\em conformal wave equation}\footnote
{note that the $R$--factor is dimension--dependent (cmp.~Wald~\cite{Wal84}, app.~D)}
$\nabla^2\:\Phi - 1/5\:R\:\Phi = 0$.
\smallskip

Transformed to an \Ef~(and asssuming $n \geq 2$), these field equations can however 
be reduced to the more familiar looking locally equivalent forms\footnote
{for notational reasons, we do not differentiate between the \Gf~$g_{ik}$ 
and the \Ef~$g'_{ik} := e^\phi\:g_{ik}$}
\bea
\label{ef_eqn}
G_{ik} &=& (n-1)/2\:\sigma\;\Theta_{ik} + \lambda\;M_{ik},  \\
\sigma\:\nabla^2\phi &=&  1/2(n-1)\;\lambda\;F^2, \\
\nabla_i\:(\lambda\:F^{ij}) &=& 0.  
\eea
Here $\lambda := e^{\phi}$, and $\Theta$ is the 
\KG~energy momentum tensor,
\be
\Theta_{ik} := \nabla_i\phi\nabla_k\phi - 1/2\;(\nabla\phi)^2\;g_{ik}.
\ee
Unfortunately, in equation (\ref{ef_eqn}) the KG--term $\Theta$ does not couple with the same factor $\lambda$
as the constant factor of the $M$--term, thus making a conventional interpretation 
in terms of a ``variable effective gravitational coupling constant'' 
$\lambda := e^{\phi}$ somewhat problematic.

The merit of the \Ef~formulation lies however in the fact,
that for this most frequently used frame, some solutions to slightly more general couplings are
already known. We will in the following take advantage of this.

\section{Static Spherically Sym\-metric Solutions}
\label{solutions}

\subsection{Smooth GDC Solutions of Magnetic Type}

In the following, we will deal exclusively with the case $n=4$ ($\sigma=1$).
The MED field equations in the \Ef~then reduce to
\bea
G_{ik} &=& 3/2\;\Theta_{ik} + \lambda\;M_{ik},  \\
\nabla^2\phi &=&  1/6\;\lambda\;F^2, \\
\nabla_i\:(\lambda\:F^{ij}) &=& 0.
\eea
We want to obtain \SSS~(SSS) 
solutions to the GDC field equations.
This is most easily done staying in the \Ef,
where SSS solutions to slightly
more general equations, depending on an extra parameter $a$,
have been found by \GHS~\cite{GHS91} (GHS).\footnote
{in the context of a broader framework, equivalent solutions
have been found earlier by Gibbons and Maeda~\cite{GiM88}
(see also Horowitz~\cite{Hor92}, relating these solutions to the GHS solution)}
They start from the following Lagrangian density,
\be
\mathcal L = |g|^{1/2}\;(R - 2\;(\nabla \phi)^2 - 1/2\;e^{-2a\phi}\;F^2),
\ee
where we trivially rescaled their $F$ with a factor~$1/\sqrt 2$
in order to have a closer correspondence with the classical Maxwell field.
Our \Gf~Lagrangian can evidently be conformally mapped to the GHS Lagrangian
with the particular choice of the parameter $a = \pm1/\sqrt 3$.

Their general ``magnetic--type'' solution can then be written as\footnote
{in accordance with the notation of GHS, $\lambda$ and $R$ here refer only to eqn.~(\ref{GHS_metric})}
\bea
\label{GHS_metric}
ds^2 &=& -\lambda^2\;dt^2 + \lambda^{-2}\;dr^2 + R^2\;d\Omega^2, \\
\lambda^2 &:=& X_+\;X_-^{(1-a^2)/(1+a^2)}, \\
R^2 &:=& r^2\;X_-^{2a^2/(1+a^2)}, \\
e^{-2a\phi} &=& X_-^{2a^2/(1+a^2)}, \\
F &=& F_m := q\;\sin{\theta}\;d\theta \wedge d\varphi, \\
\mbox{where} \quad X_+ &:=& 1 - r_+/r, \quad X_- := 1 - r_-/r.
\eea
The parameters $q,\:r_+,\:r_-$  are however restricted by\footnote
{despite the suggestive notation, it is {\em not} required that $r_+ \geq r_-$}
\be
q^2 = 2\;r_+r_- /(1+a^2).
\ee

Mapping the $a^2=1/3$ GHS--solution back to the \Gf, 
we then get the following ``magnetic'' family of solutions,
\be 
ds^2 = - X_+\;dt^2 + (X_+ X_-)^{-1} dr^2 + r^2 d\Omega^2, 
\ee
\be
e^{\phi} = X_-^{1/2}, \quad F = F_m := q\;\sin{\theta}\;d\theta \wedge d\varphi, 
\ee 
\be
\mtext{where} \quad q^2 := 3/2\;r_+r_-.
\ee

However, in these coordinates the metric is {\em not} regular at $r=r_-$, 
and the dilaton scalar does not even exist for $r < r_-$.
Introducing a new coordinate $\rho$ 
(solution of $d\rho/dr = \varepsilon\;X_-^{-1/2}$) by means of
\be 
\rho(r) = \varepsilon \;r_-\;\left( \xi^{1/2}(\xi-1)^{1/2}
+ \ln \left(\xi^{1/2} + (\xi-1)^{1/2} \right) \right), \qquad \xi := r/r_-,
\ee 
removes these drawbacks.
The parameter $\varepsilon = \pm 1$ characterizes the two \mbox{$\rho$--branches}
joined at $\rho = 0$ and mapping to the single $r$--range $r \geq r_- > 0$.
The inverse function $r(\rho)$ is smooth in $\rho$.
In these new cordinates (which now properly include the locus $r=r_-$,
resp.~$\rho=0$),
the resulting metric 
\be 
ds^2 = -X_+ \; dt^2 + X_+^{-1} \; d\rho^2 + r^2(\rho)\;d\Omega^2, 
\ee 
is nondegenerate\footnote
{except for the usual easily removable degeneracy at the $z$--axis} 
and can be shown to  be {\em smooth} in a neighbourhood of $\rho=0$
which does not include $\rho=\rho(r_+)$.\footnote
{note the close similarity of the form of this metric 
with Schwarzschild's, to which it reduces for $r_-=0$}
Also, $e^{\phi}$ becomes smooth there,
and the expression for the Maxwell field $F_m$ remains unchanged
(and smooth).

As the metric is symmetric under $\rho \to -\rho$, 
and there is now always a two--sphere with minimal--area $A=4\pi r_-^2$ 
(corresponding to a ``radius'' $r(0)=r_-$),
the geometric interpretation is that of a {\em wormhole}
with throat at $\rho=0$.
This notion of wormhole, based on a {\em local reflection symmetry}, 
is however different and more general than the usual one,
which only allows wormholes with {\em timelike} throats
(e.g.~Visser~\cite{Vis95}, cmp.~also Hayward~\cite{Hay02}).

If $r_+ \geq r_-$ the locus $r=r_+$ can be shown to be a {\em regular event horizon}
(even in the ``degenerate'' case $r_+=r_-$),
and the metric can be smoothly extended through it by standard procedures
(e.g.~by using \EF~coordinates).
For $r_+ < r_-$ there is no black hole
and the complete metric can be expressed in the single coordinate chart given above.
The wormhole topology (connecting universes with their own asymptotic regions)
is common to all metrics of this family and can in this sense be considered 
as {\em generic} among the class of SSS solutions.
\medskip

The Carter--Penrose (CP) diagrams corresponding to these extensions 
fall into the three distinct types I: $r_- > r_+ \geq 0$, 
II: $r_- = r_+ > 0$, III: $r_+ > r_- > 0$ 
and are shown in the accompanying figure 
(where we also used the more physical characterization by means of
charge $q$ vs.~mass $m$, to be justified later).
Depending on the type of the solution, the throat of the wormhole is {\em timelike}
for type~I, {\em null} for type~II (coinciding with the event horizon)
and {\em spacelike} for type~III.

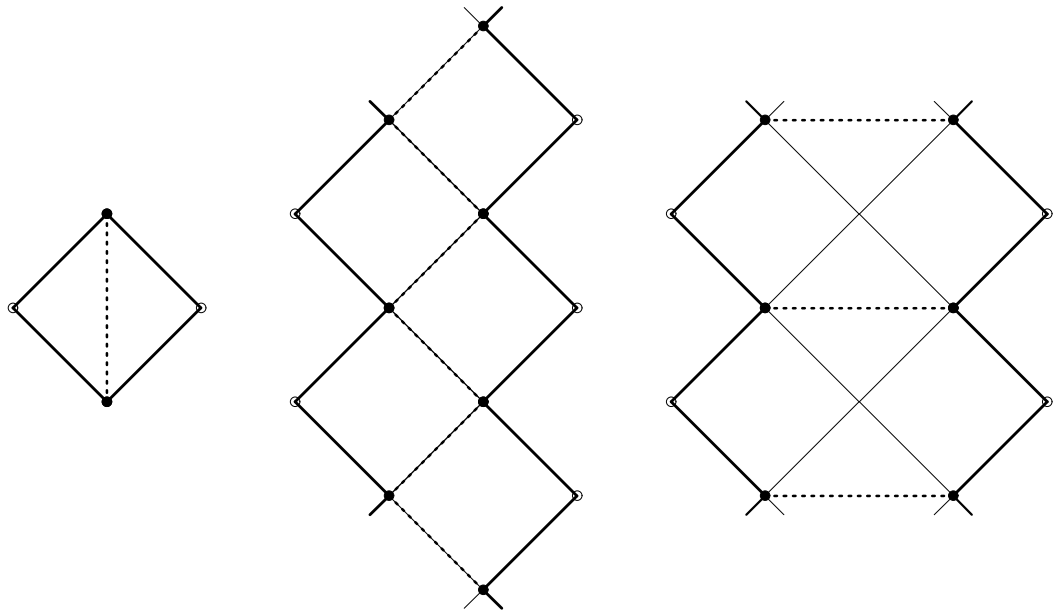
\newsavebox{\BHRegion}
\savebox{\BHRegion}
{
\setlength{\unitlength}{0.25cm}

\begin{picture}(10,10)

\thinlines
\drawline(0,5)(5,10)
\drawline(0,5)(5, 0)

\Thicklines
\drawline(10,5)(5,10)
\drawline(10,5)(5, 0)

\thinlines
\put( 5, 0){\circle*{0.5}}
\put( 5,10){\circle*{0.5}}
\put(10, 5){\circle {0.5}}

\end{picture}
}

\newsavebox{\HBRegion}
\savebox{\HBRegion}
{
\setlength{\unitlength}{0.25cm}

\begin{picture}(10,10)

\Thicklines
\drawline(0,5)(5,10)
\drawline(0,5)(5, 0)

\thinlines
\drawline(10,5)(5,10)
\drawline(10,5)(5, 0)

\put( 5, 0){\circle*{0.5}}
\put( 5,10){\circle*{0.5}}
\put( 0, 5){\circle {0.5}}

\end{picture}
}

\begin{figure}[h]

\setlength{\unitlength}{0.25cm}
\begin{picture}(50,43)(4,-8)
\put(0,10){\usebox{\BHRegion}}
\put(0,10){\usebox{\HBRegion}}
\thicklines
\dottedline{0.5}(6.1,10)(6.1,20)

\put(15,15){\usebox{\HBRegion}}
\put(15,05){\usebox{\HBRegion}}
\put(20,10){\usebox{\BHRegion}}
\put(20,20){\usebox{\BHRegion}}
\put(20,00){\usebox{\BHRegion}}

\thinlines
\drawline(26.1,0)(25.1,-1)
\drawline(26.1,30)(25.1,31)
\Thicklines
\drawline(26.1, 0)(27.1, -1)
\drawline(21.1, 5)(20.1,  4)
\drawline(21.1,25)(20.1, 26)
\drawline(26.1,30)(27.1, 31)

\thicklines
\dottedline{0.5}(26.1, 0)(21.1, 5)(26.1,10)(21.1,15)(26.1,20)(21.1,25)(26.1,30)

\put(35,15){\usebox{\HBRegion}}
\put(35, 5){\usebox{\HBRegion}}
\put(45,15){\usebox{\BHRegion}}
\put(45,05){\usebox{\BHRegion}}

\thicklines
\dottedline{0.5}(41.1, 5)(51.1, 5)
\dottedline{0.5}(41.1,15)(51.1,15)
\dottedline{0.5}(41.1,25)(51.1,25)

\thinlines
\drawline(41.1, 5)(42.1, 4)
\drawline(51.1, 5)(50.1, 4)
\drawline(41.1,25)(42.1,26)
\drawline(51.1,25)(50.1,26)

\thicklines
\drawline(41.1, 5)(40.1, 4)
\drawline(51.1, 5)(52.1, 4)
\drawline(41.1,25)(40.1,26)
\drawline(51.1,25)(52.1,26)

\put( 0,-5){{\bf Type I: $r_- > r_+ \geq 0$}}
\put(18,-5){{\bf Type II: $r_- = r_+ > 0$}}
\put(40,-5){{\bf Type III: $r_+ > r_- > 0$}}

\put(1.0,-7){{\bf $(q^2 > 6\;m^2 > 0$}, and}
\put(1.0,-9){{\bf $q = m = 0,\;r_- > 0)$}}
\put(20,-7){{\bf $(q^2 = 6\;m^2 > 0)$}}
\put(43,-7){{\bf $(0 < q^2 < 6\;m^2)$}}

\end{picture}

\caption{Carter--Penrose diagrams for the extended Wormhole Metrics.\\
Thick lines: null infinity, thin lines: event horizon, dashed lines: wormhole throat, circles: $i^\pm,\:i^0$}

\end{figure}

\subsection{Gauge Potential for the Magnetic GDC Solution}

Although the question of an appropriate gauge potential
is most often ignored (being trivial in the ``electric'' case), 
we will now exhibit a smooth potential
in the sense of an $U(1)$--gauge theory. 
The existence of such a potential makes the smooth SSS solution complete.
Consider the $u(1)$--valued (i.e.~purely imaginary)
\be
\tilde A_{\pm} := -i\:n/2\;(\cos \vartheta \mp 1) \: d\varphi,
\ee
where the upper sign refers to the upper hemisphere $\vartheta \neq \pi$
and the lower sign to the lower hemisphere $\vartheta \neq 0$.
Evidently, $\tilde F := d\tilde A = i\:n/2\;\sin \vartheta\;d\vartheta \wedge d\varphi$.
The transition function for the potential in the overlap of the two hemispheres is 
given by $S := e^{in\varphi}$: $A_+ = A_- + S^{-1}dS$. 
For consistency $n \in \mathbf N$ must hold (cmp.~G\"ockeler and Sch\"ucker~\cite{GoS87}
for more details).
Reverting to the corresponding real field, $i\:F := \tilde F$,
this  amounts to $q = n/2$.
Taking properly into account the terms appearing in the ``gauge derivative''
for an electrically charged particle in the field of a magnetic monopole, 
$\nabla = \partial + ie/\hbar\:A$,
we obtain {\em Dirac's quantization condition:} $pq/\hbar = n/2$.
For $n=1$ and $p=e$ the minimal magnetic charge is
$g = 1/2\;e/\hbar \sim 68.5\:e$, 
giving the factor $(g/e)^2 \sim 4.7\times 10^3$ needed in section~\ref{interpretation}.

\subsection{Singular GDC Solutions of Electric Type}

As already shown by \GHS \cite{GHS91}, from a magnetically charged solution
$(g,\;\phi,\;F)$ of their \Ef~field equations, an electrically charged
one can formally be obtained by taking $(g,\;-\phi,\;*F)$,
where $*F$ is the (generalized) Hodge--dual of $F$.\footnote
{a manifestation of the ``weak/strong coupling duality'' of String Theory}
But then, the transformation back to the \Gf~inevitably leads to a metric 
immanently degenerate at $r=r_-$
\be
ds^2 = X_-\;\Big(-X_+\;dt^2 + X_+^{-1}\;d\rho^2 + r^2(\rho)\;d\Omega^2\Big),
\ee
which is the image of the degenerate conformal mapping with factor $X_-$
of the smooth ``magnetic'' metric considered before.\footnote
{note that $X_-$ is nonnegative, considered as function of $\rho$}
Dilaton factor and Maxwell field are given by
\be
e^\phi = X_-^{-1/2}, \qquad F = F_e := p/r^2\;X_-^{1/2}\;d\rho \wedge dt,
\ee
where $p^2 = 3/2\:r_+r_-$.
A  gauge potential for $F$ is 
\be
A_e := \varepsilon\:p/r_-\:X_-\:dt.
\ee
Note that no charge quantization is involved and that $F$ vanishes
at the singularity of the metric, $\rho=0$ ($r=r_-$) ---
in fact, both $F$ and its potential are smooth there.

If $r_+ > r_-$ and $r_+ > 0$, the locus $r=r_+$ is a regular event horizon, 
hiding the spacelike singularity. 
For the ``degenerate'' case $r_+ = r_-$, there is still a horizon,
but the singularity becomes timelike. 
For $r_+ < r_-$, the singularity is timelike and even naked.
The corresponding CP diagrams agree
with those of the \RN~family of solutions,
except for the case $r_+ > r_-$, where the diagram agrees with \Sch's,
which has a spacelike singularity.

\subsection{Alternative Dilaton Coupling} 

A closer look at the general GHS solution reveals
that only the choice $a^2=1/3$ allows to
remove the offending common $X_-$--factor from the SSS metric by 
an appropriate conformal transformation.
This again is given by the same factor $e^{-2a\varphi}$,
as from the \Gf~to the \Ef,
resulting in the ``alternative'' ADC Lagrangian
\be
\mathcal L = |g|^{1/2}\:(e^{-\phi}\:R - 1/2\:e^{\phi}\:F^2).
\ee
Evidently, it {\em cannot} directly be interpreted in terms
of a volume manifold and corresponding coupling scheme.
The field equations corresponding to this alternative Lagrangian are then
\bea
G_{ik} - \Theta'_{ik} + \Theta''_{ik} &=& e^{2\phi}\:M_{ik}  \\
R &=& -1/2\;e^{2\phi}\:F^2  \\
0 &=& \mbox{div}\;(e^{2\phi}\:F), 
\eea
with corresponding dilaton equation derived from them,
\be
\square \; \phi = -1/6\; e^{2\phi}\:F^2.
\ee
Here the divergence and the Laplacian are defined
based on a volume element density $|g|^{1/2}\:e^{-\phi}$.
Except for the manifest appearance of dilaton factors
the essential change is a sign reversal in the dilaton equation.

This shows that smooth electric wormhole solutions are
possible, when sacrificing the geometrical interpretation.
Their metrics agree with those of the smooth GDC wormholes.
The coupling is again ghost--free and the material part
still obeys the energy conditions.
Among the class of MED Lagrangians considererd,
it is the only Lagrangian with smooth SSS solutions.

An ``alternative'' coupling scheme, involving arbitrary nongeometrical fields, 
could however be formulated tentatively as follows:\\
i) apply the standard GDC scheme, 
ii) denote the map from the \Gf~to the \Ef~by $\mathcal E$ and
iii) define the ``alternate''  \Af~(and  corresponding Lagrangian)
as the image of the \Gf~under the map $\mathcal E^2$.

When including point masses, such a coupling scheme
would lead to nongeodesic behaviour for their trajectories
(see section~\ref{interpretation}).
This could be considered as a drawback.
But the main objection against this coupling is of course
that it has been deliberately constructed so to possess
smooth electrically charged SSS solutions,
and also its lack of any direct geometric interpretation.

\subsection{Comparison with other SSS Solutions}

As already noted by \GHS,
{\em all} the nontrivial \Ef~metrics of the GHS family 
of solutions are either geodesically incomplete and/or singular,
with the exception of the ``cornucopion'' metric,
which is an extreme solution for $a=1$ interpreted in the string--frame.

To my knowledge all other 
\SSS~solutions directly or indirectly
related to \MED~gravity violate some of the energy conditions,
and must be considered as classically ``unphysical''.\footnote
{in fact, Bronnikov's wormhole solutions turn out to be highly unstable (cmp.~\cite{BrG02}).
This is also the case for the recently found ``ghostly'' massless wormhole solution 
of \mbox{\ArPi}~\cite{Arm02}, as discussed by Shinkai and Hayward~\cite{ShH02}}
Therefore we will not dicuss them here.
\medskip

Unfortunately, 
general enough existence or no--go theorems do not yet exist,
save for particular couplings and the corresponding conformal frames.
For example, for the closely related vacuum Brans--Dicke theory, it has been shown by
Nandi, Bhattacharjee, Alam and Evans~\cite{NBA98} that while in the Jordan--frame
there do exist wormhole--solutions for the (unphysical) range $-3/2 < \omega < -4/3$
of the BD--parameter, which are however plagued by ``badly diseased'' naked singularities, 
in the \mbox{\Ef} there do not exist such solutions at all, unless energy--violating
regions are deliberately introduced.
\medskip

However we must admit that while our GDC/ADC wormhole solutions are
smooth as regards metric, volume element density (i.e.~dilaton--factor $e^\phi$, resp.~$e^{-\phi}$)
and gauge potentials, they are not, when instead considering
the dilaton scalar $\phi$ itself, which diverges to $-\infty$
(resp.~$+\infty$) at the throat of the wormhole.
Although this poses no problem for the smoothness of
the corresponding dilaton factors (which just smoothly vanish there),
the geometric interpretation in terms of a 
corresponding nondegenerate volume element density
cannot anymore strictly maintained,
as it is not  manifestly positive.
This must in fact considered to be a flaw, albeit not a serious one,
as the physical interpretation does not suffer.

\section{Interpretation}
\label{interpretation}

\subsection{GDC/ADC and Einstein's Equivalence Principle}

Let us  emphasize that the GDC theory presented here
is still a {\em metric} theory, in the sense that massive point
sources move on timelike {\em geodesics} of the metric,
when expressed in a \Gf.
This can be most easily seen by noting that the volume integral over
the GDC Lagrangian for a point particle
effectively splits into the volume integral over the ``geometric''
part of the Lagrangian plus a conventional line integral
$m \int (\dot x^2)^{1/2}\;d\tau$,
which by variation of the arc length then leads to the standard
geodesic equation.\footnote
{this can also be justified with a ``pure dust'' model of matter,
based on relativistic thermodynamics (paper in preparation)}

As the alternative coupling scheme seems to be
too artificial to be really believed,
we introduce the point particle term
``by hand'' into the \Af~Lagrangian.
Therefore Einstein's Equivalence Principle (EEP) is satisfied
both for the GDC and (trivially) ADC formulations.

However, for both the standard GHS and the generalized GHS solution
the validity of the EEP is undecided,
as an equivalent to a coupling scheme to external sources has not been formulated.
In the following we will therefore simply {\em assume} the EEP
to hold also for the corresponding string--inspired theory.
This will allow us in the following to interpret the PPN parameter $m$ 
for {\em all} solutions considered
as the mass of the gravitational source.

\subsection{PPN Viability and
Experimental Verifyability}

However, the compatibility with the EEP is not sufficient for the viability
of a generalized theory of gravity, as it does not refer to any particular solution.
A framework to check just this (in particular, SSS solutions)
is well--known under the name of Parametrized
Post Newtonian (PPN) Approximation (cmp.~Will~\cite{Wil81}).
We will give in tabular form only the resulting most important parameters
mass $m$, $\beta$ and $\gamma$,
and compare them on the one hand with the corresponding parameters
for the $\alpha^2=1$ GHS solutions (parametrized by $(M,Q,\phi_0)$,\footnote
{$M$ is not necessarily to be identified with the PPN mass}
where $M:=r_+/2,\;q^2:=r_+r_-$,\footnote
{note that we have to rescale $F$ with $1/\sqrt2$ in order to have common conventions}
$q := Q\;e^{-\phi_0}$), 
both in the \Ef~and in the \Sf,
and on the other hand with the PPN parameters of the \RN~metric, also given in the same form\footnote
{note that this form of the metric is only possible, if $m^2 \geq q^2$, i.e.~$\epsilon^2 \leq 1$}
\be
ds^2 = -\lambda^2\;dt^2 + \lambda^{-2}\;dr^2 + r^2\;d\Omega^2, \mtext{where} \lambda^2 := (1-r_+/r)(1-r_-/r).
\ee

In this PPN approximation, the metric has the following ``isotropic'' form
\be
ds^2 = -(1 - 2\:m/r + 2\:(m/r)^2 \: \beta)\:dt^2 + (1 + 2\:m/r \: \gamma)(dr^2 + r^2 \: d\Omega^2).
\ee
Setting $\epsilon := q/m$, $x := r_-/r_+$
and assuming $m \neq 0$, $r_+ \neq 0$, 
we can collect the results in the following table,
where also the asymptotic behaviour of $\gamma$ is given for $\epsilon \to \infty$.
As for the electric GDC solution $\epsilon$ is limited by $\epsilon^2 \leq 3/2$,
in this case an asymptotic expression does not make sense.
The metrics for the GDC and ADC wormholes agree, so they share a common table entry.

\begin{table}[h]

\begin{tabular}{lllll}
\hline \hline
\rule[-2mm]{0mm}{2mm} & \ \ $2\:m$ & \ \ $\beta$ & \ \ $~\gamma$ & \ \ $\gamma$--asymptotics \\
\hline 
GDC/magnetic \rule[2mm]{0mm}{3mm} & & & & \\
ADC/electric & \ \ \raisebox{1.5ex}[-1.5ex]{$r_+$} & \ \ \raisebox{1.5ex}[-1.5ex]{$1 + 1/6\;\epsilon^2 \ $} 
& \ \ \raisebox{1.5ex}[-1.5ex]{$~\:\!1 + 1/6\;\epsilon^2 \ $} & \ \ \raisebox{1.5ex}[-1.5ex]{$O(\epsilon^2 \ )$}  \\
GDC/electric & \ \ $r_++r_-$ & \ \ $1 + 1/3\;\epsilon^2 \ $  & \ \ $(1+x)^{-1}$ & \ \ n.~a.  \\
GHS/\Ef ~ & \ \ $r_+$ & \ \ $1 + 1/4\;\epsilon^2 \ $ & \ \ $~\:\!1$ & \ \ $O(\epsilon^0)$  \\
GHS/\Sf ~ & \ \ $r_+-r_-$ & \ \ $1$ & \ \ $(1 + x)/(1 - x)$ & \ \ $O(\epsilon^1)$ \\
\RN~\rule[-3mm]{0mm}{3mm} & \ \ $r_++r_-$ & \ \ $1 - 1/2\;\epsilon^2 \ $ & \ \ $~\:\!1$ & \ \ $O(\epsilon^0)$ \\
\hline \hline
\end{tabular}

\caption{PPN parameters for charged \SSS~metrics}

\end{table}

Before proceeding, let us observe that the \Sf~GHS ``cornucopion'' solution, 
being degenerate ($r_+=r_-$), has {\em vanishing} PPN mass $m$ 
and so does not fit into the table
($\beta$ and $\gamma$ undefined).
This applies also to both the zero--mass magnetic GDC 
and electric ADC solutions, where $r_+=0$.

For heavenly bodies,
where most empirical data comes from,
we have $|\epsilon| = |q/m| \ll 1$.
Assuming $r_+ \approx m$,
we also have $x := r_-/r_+ \equiv 2/3\:(q/r_+)^2 \approx \epsilon^2 \:(m/r_+)^2 \ll 1$,
so all these metrics are {\em equally viable},
having $\beta \approx \gamma \approx 1$.

However, applied to charged elementary particles
(where up to now no PPN data seems to exist),
the corresponding $\gamma$ show significant differences in
their dependence on the ``specific charge'' ratio $q/m$.
This is most pronounced for the smooth SSS
metric, where $\gamma \sim \epsilon^2 \ $.
For the smooth electric metric, taking the charge and mass
of the electron, we get the factor $\epsilon^2 \approx 10^{40}$,
which is essentially the smallest of Dirac's Big Numbers.

This would lead for the electric ADC metric to signi\-ficant
effects near the wormhole throat $r \approx r_- \approx q^2/m$,
which for the electron is of the order of the 
{\em classical electron radius} $r_{_{\!El}} \approx 3\times10^{-13}$~cm.
For the magnetic GDC metric, this estimate does change less than one order of magnitude,
when taking $q = e/2\alpha$ (corresponding the Dirac's charge quantization) 
and assuming the mass of the proton, $m \sim 1836\:m_e$,
giving a throat ``radius'' of about $8\times10^{-13}$~cm.
The energies corresponding to these distances 
(a few MeV) would be well within the reach of present--day
experimental technology.

In comparison, for the \Sf~GHS metric deviations would appear 
only at distances (and corresponding energies) of the order
of $r \approx e$, which is even one order of magnitude smaller 
than the Planck length $r_{_{\!Pl}} \approx 2\times10^{-33}$~cm.

Of course, magnetic monopoles have not been observed
(not even by their characteristic electromagnetic signature).
However for the possible dilatonic effects of electrically charged particles
the situation looks more favourable.
A realistic estimate of the effects involved would have to take into account
a (not yet existing) Quantum Field Theory adapted to curved backgrounds
with nontrivial topology.

\subsection{Zero--Mass Solutions; Boundedness of Mass}

From a formal geometric viewpoint, the classical \RN~family of metrics 
are equally meaningful for zero and even negative values of their mass parameters ---
only the character of their ever--present singularities at $r=0$
change to the worse.
From a physical standpoint, 
one would like to be able to exclude such negative--mass solutions.
In the classical Einstein theory for isolated systems, 
and assuming suitable energy conditions,
this has been achieved only fairly recently
(cmp.~references given in Wald~\cite{Wal84}).

The boundedness of the mass, $m \geq 0$, is however guaranteed by the 
families of charged SSS  solutions described above, 
when insisting on {\em smoothness}.

Firstly, the particular solutions with $r_+=0,\;r_- > 0$ have smooth 
metrics with {\em vanishing mass}, $m=0$.
As the corresponding charge $q$  necessarily vanishes,
they are {\em vacuum solutions} of GDC/ADC gravity.\footnote
{closely related uncharged massless wormhole solutions have been
investigated very recently by \ArPi~\cite{Arm02}, 
although for a ``ghostly'' KG/dilaton scalar}

Now, let us assume negative mass $m$, i.e.~$r_+ < 0$.
Then we must necessarily also have $r_- < 0$,
if the solution is to be charged.
But then the transformation $r \mapsto \rho(r)$
to regular coordinates fails to produce a metric
locally regular at $\rho(0)$ --- 
regularity can also not be achieved by any other map.
Remain the uncharged possible solutions with $m < 0$.
This means necessarily $r_-=0$ --- but this is
exactly the negative--mass \Sch~metric,
which is well--known to have a naked singularity.\footnote
{the positive--mass \Sch~metrics
would also be excluded by smoothness} 
Therefore as claimed, for the family of smooth SSS solutions
there must be the {\em lower mass bound }
$m \geq 0$, in order that the metric remains smooth.
The nonflat smooth SSS metrics can thus for $m \neq 0$  be characterized
by the two physically meaningful parameters $m,\;q$, 
where  $m > 0,\;q \neq 0$, whereas the massless solutions ($m=0$) 
are  uncharged ($q=0$) and are 
uniquely characterized by their ``scalar charge'' $r_-\geq0$.

Incidentally, the conserved Total Energy residing in the Maxwell field
can be straightforwardly calculated, giving $E=3\;m$ for both types~I and~II
and $E=q^2/m < 3\;m$ for type~III, indicating that there is
a {\em gravitational binding energy}  $\Delta_g E \leq 2\;m$,
saturated for type~I and~II.
Note that there is consistency in the sense that 
the vanishing of the mass $m$ corresponds to the 
vanishing both of the charge $q$ and the Maxwell field $F$.

\subsection{Energy Conditions; Repulsion}

Due to fairly general theorems, in the context of classical gravity
(e.g.~Friedman, Schleich and Witt~\cite{FSW93},
cmp.~also Visser's monography~\cite{Vis95}), 
wormhole metrics like the smooth SSS metrics
would necessarily somewhere exhibit ``exotic matter'',
in the form of regions with negative energy density
of the material source.\footnote
{the few known ``nonexotic'' wormhole solutions of the classical Maxwell--Einstein theory
(e.g.~Schein and Aichelburg~\cite{ScA96}) in fact break some of the standard assumptions,
like  having closed timelike lines}
The Maxwell stress tensor $M_{ik}$ evidently obeys 
automatically even the Dominant Energy Condition (DEC)
and was used as the only material source in our
system of field equations --- moreover it was
coupled in the ``orthodox'' way (up to a dilaton factor in the case of the electric ADC solution)
on the r.h.s.~of the field equation.
$G - \Theta' - \Theta'' = M$, resp.~$G - \Theta' + \Theta'' = e^{2\phi}\:M$.
Therefore it can be said that our solutions do {\em not}
contain any exotic matter, their material sources
obeying the DEC.
This is also evident in the \Ef, where the 
KG stress tensor (adding to the Maxwell 
stress tensor multiplied by the dilaton factor) obeys the DEC.
In fact, the energy tensor for our dilaton metrics does even obey the Strong Energy Condition
(SEC), which plays a prominent role in the singularity theorems of classical gravity.

So how it comes that there seems to be a ``repulsive force'' holding open the throat of the wormhole?
This can be seen by invoking the \Ray~identity for geodesics.
First note that this identity cannot be applied directly in the \Gf,
as the field equations involve extra geometric terms besides the Einstein tensor.
And mapping  geodesics from the \Gf~or from the \Af~to the \Ef~(where Einstein's equations formally hold) 
will result in additional nongeodesic terms: 
$\dot {\vec u} = 0 \to \dot {\vec u} = 1/2\;((u\cdot d\phi)\;\vec u + g^{-1}d\phi)$.%\footnote
%{for timelike geodesics in the \Gf, and up to rescaling and reparametrization in the \Ef}
Such terms $\sim d\phi$ are also known from Nordstr\"om's scalar theory of gravitation
(a precursor of Einstein's metric theory).\footnote
{for a short history of scalar--tensor theories of gravitation, see Brans~\cite{Bra97}}
These additional terms also prevent the \Ray~identity to be applied.
For the smooth SSS metrics (and $r > 2m$) they have a repulsive effect, 
$g^{-1}d\phi=1/2\;r_-/r^2\;(X_+/X_-)^{1/2}\;n_\rho$ ($n_\rho := X_+^{1/2}\;\partial_\rho$),
which for type~I and type~III becomes unbounded at the former locus of the throat,
where the metric is now singular.
Therefore, when interpreted in the \Ef, test particles seem to be always effectively {\em repelled}
by the object ``sitting'' at $\rho=0$.
In contrast, being driven by $e^{-\phi}$,
the singular electric SSS GDC solution would always appear to attract the test particle.

\section{Conclusions}
\label{conclusions}

With respect to the criteria mentioned in the introduction,
we have been able to show that among the class of \MED~Lagrangians
there exist two essentially different couplings allowing for
well--behaved \SSS~solutions.
However, only the GDC Lagrangian admits a simple coupling scheme.
Moreover, it has an immediate geometric interpretation
in terms of a Volume Manifold.
\smallskip

When it comes to make a choice between alternative
Lagrangian--based theories of dilaton gravity,
of course it depends on the weight one is willing to give to the
existence of well--behaved (i.e., smooth) solutions
and to the generality of MED Lagrangians.
In the context of classical gravity, the ``regularizing'' nature of the GDC/ADC dilaton 
could be welcomed as a ``new degree of freedom'' to tame
some of the inherent divergences.
The main obstacle is however still the ``magnetic'' nature of the geometric GDC solutions,
whereas for the ``electric'' ADC solutions, it is their lack of geometric interpretation
and the non--uniqueness of a corresponding coupling scheme.
There are no indications that more ``sophisticated'' Lagrangians
could resolve this dilemma.
\medskip

The stability of the smooth wormhole solutions has not been touched
in our work, and constitutes the major open issue.
However, in view of the fulfillment of all the energy conditions,
the prospects seem to be promising.
\smallskip

It would be also be highly desirable to have general {\em stationary}
spherically symmetric GDC/ADC solutions. 
Some basic questions in this context: 
are there still smooth charged wormhole solutions? 
Do nontrivial geometric  vacuum solutions again exist? 
Does the ``magnetic--electric dilemma'' still persist? 
Can stability proven to hold in a general sense?

A satisfactory answer to these questions would of 
course challenge the role of Classical Relativity 
and the corresponding (nondilatonic) Black Holes.
\smallskip

Applications to the different setting of cosmology
would be particularly interesting: there is the possibility that the dilaton
could again act repulsively, thus contributing to the observed accelerated cosmic expansion.
\smallskip

The diverse global spacetime models derivable
from the basic extensions described here could also serve as ``sandboxes'' 
to develop and test some other fundamental theories 
in situations where the two--dimensional formulations are too limited
to be realistic.

\begin{acknowledgments}
I want to express my gratitude to Peter Aichelburg for useful suggestions.
After submitting this paper for publication, 
I learned that a MED Lagrangian essentially equivalent 
to ours was already postulated by Cadoni and Mignemi \cite{CaM95},
with the primary aim of having a 4D generalization of the 2D 
Jackiw--Teitelboim theory.\footnote
{for a recent review of 2D dilaton gravity, see Grumiller, 
Kummer and Vassilevich \cite{GKV02}} 
Although they emphasized the 2D aspects,
they also noted that the corresponding 4D field equations admit 
nonsingular (but apparently geodesically incomplete) magnetically 
charged black hole solutions.

\end{acknowledgments}

\end{document}